# Influence of pH and sequence in peptide aggregation via molecular simulation

Marta Enciso[*]

*Department of Chemistry and Physics,*
*La Trobe Institute for Molecular Science, La Trobe University, Australia*

Christof Schütte[†]

*Institute for Mathematics, Freie Universität Berlin, Germany and*
*Zuse Institute Berlin, Germany*

Luigi Delle Site[‡]

*Institute for Mathematics, Freie Universität Berlin, Germany*

## Abstract

We employ a recently developed coarse-grained model for peptides and proteins where the effect of pH is automatically included. We explore the effect of pH in the aggregation process of the amyloidogenic peptide KTVIIE and two related sequences, using three different pH environments. Simulations using large systems (24 peptides chains per box) allow us to correctly account for the formation of realistic peptide aggregates. We evaluate the thermodynamic and kinetic implications of changes in sequence and pH upon peptide aggregation, and we discuss how a minimalistic coarse-grained model can account for these details.

---

[*]Electronic address: `m.enciso@latrobe.edu.au`
[†]Electronic address: `schuette@zib.de`
[‡]Electronic address: `luigi.dellesite@fu-berlin.de`

## I. INTRODUCTION

Neurodegenerative diseases and protein-based drugs have put protein aggregation in the limelight, drawing a large research interest over the last decades [1–3]. In fact, nonspecific aggregates and fibrils have been identified as a core element in many degenerative diseases and dementias [4]; peptide drugs such as new antibiotics need to present enhanced stability and resistance upon aggregation to be competitive as commercial drugs [5].

In spite of the large efforts made in the field, aggregates are still hard to characterize with microscopic detail using standard protocols; their amorphous state makes them impractical for the usual high resolution techniques such as nuclear magnetic resonance and X-ray crystallography. Among the alternative strategies that are commonly used to investigate protein aggregation, computer simulations play a very significant (and increasingly important) role. First, simulation provides a microscopic description of the system (according to the assumptions of the simulation model). Second, it gives an optimal framework to test sequences, structures and environment conditions in a systematic way. As a result, simulations can aid in the design of experiments and molecules tailored for properties on demand [6].

However, aggregation processes are also a major challenge for the computational community due to the interplay of many physical scales (in space and time). In particular one needs to bridge atomistic detail at time scales of picoseconds (typical of interatomic interactions) to the macroscopic scale of multipeptide systems that aggregate on the millisecond scale. As a result, most atomistic simulation studies either focus on just a small part of the whole process [7] or a minimal system size [8, 9]. Coarse-grained models are, then, an appealing alternative. Thanks to their simplified description of the system, they allow the study of larger and more complex systems while keeping a reasonable computational cost [10, 11].

Protein aggregation is a ubiquitous event that may happen to any known protein [12, 13]. Aggregation propensities, however, depend on the particularities of protein sequence [14] and structure [15], as well as on external factors like pH, concentration, ionic strength, to cite a few. Among the latter, pH is of key importance because of its clinical relevance (it has even been suggested as a diagnosis method [16]) and its impact in drug production and synthesis [17]. From a computational point of view, media acidity is also a challenge, as a proper description of the system pH implies instantaneous changes in the protonation states of the involved species. Some simulation strategies have been proposed during the last decades, all of them focused on atomistic descriptions of the biological systems [18, 19]. The most relevant drawback of constant-pH atomistic simulations for aggregation studies is its computational cost (they are remarkably slower than standard simulations), which usually makes these models impractical for aggregation purposes. Coarse-grained pH-dependent methods, in contrast, are still competitive and, if a discrete protonation approach is used, do not even have a large impact on computational efficiency.

We have recently proposed a constant-pH simulation algorithm in combination with a simple but accurate coarse-grained force field [20, 21]. It provides satisfactory qualitative results in aggregation studies when compared to both experimental and full-atomistic simulation data. Our strategy is based on the principle of "consistency across scales", ie. microscopic data taken from atomistic simulations and experiments is used to build a first version of the coarse-grained model, which is then employed to study a system at different conditions from those used as input. The coarse-grained simulation results are then compared with atomistic simulations and/or experimental data obtained at the modified conditions. The coarse-grained model is refined until both sets of results agree; then, the model is accepted and can be employed in larger simulations (see Ref.[22] and references therein for the basic strategy). This technique assures a certain universality to the model and carries the essential microscopic information for the aggregation process; as a result, qualitative results are indeed satisfactory, whereas quantitative information may differ from higher resolution techniques, as the specific chemistry of the system is unavoidably simplified.

It was mentioned above that aggregation is a common feature of every protein [14]. This fact allows the use of test systems to explore its fundamental characteristics. Test systems are also very convenient from a computational point of view because they reduce the complexity of the system. It is easier, then, to break the overall process into its individual factors, extracting more powerful and clear conclusions. In this particular study we have chosen the *de novo* sequence KTVIIE as test system, designed by López de la Paz and co-workers to form amyloids [23, 24]. Literature is rather broad regarding this family of peptides and hence it provides an broad range of data to use for comparison in a computational study like the current one. In particular, we have analyzed the effect of pH in aggregation; moreover the "amyloidogenicity" of different sequences has been investigated. Among the works on these and closely related peptides [25–27], only one has taken the effect of pH into account [27]. The limitations of atomistic constant-pH molecular dynamics simulations make this study rather constrained in terms of the system size, ie. formation of a dimer in a 4-amino acid peptide. That work, although very valuable, can just extract limited conclusions about peptide aggregation, as the studied system is smaller than the critical nucleus size of aggregating peptides, usually set between four and ten peptides [11, 28–30]. In the present study, we use Replica Exchange Monte Carlo simulations to extract thermodynamic information and a Kinetic Monte Carlo for qualitative dynamic data. We use these computationally efficient simulation techniques jointly with a recently reported constant-pH coarse-grained strategy [20, 21]. Their combination has allowed us to increase by an order of magnitude the systems' size, as well as the exploration of several pH conditions and sequence alternatives within the same study. In this way, our results can be realistically compared with macroscopic experimental information.

## II. METHODS

In this section we detail the technical aspects regarding this work, particularly the pH-dependent coarse-grained methodology as well as some notes about the simulation techniques that we used: Replica Exchange Monte Carlo (REMC) and Kinetic Monte Carlo (KMC). They have been described in full detail somewhere else and just a brief description is included here (see references in the following subsections).

### A. System description and force field

Peptides are described by an off-lattice bead and stick model. Each amino acid is described by just one interaction center, placed at the $\alpha$-carbon position; the molecule is then embedded in an environment described via implicit solvent approximation. Neighboring beads within the same peptide chain are located at a fixed distance of 0.38 nm (i.e. the average length of a *trans* peptide bond).

The central assumption of the model, based on past experience and success, is that the main peptide driving forces are considered independent additive terms in the overall force field [10, 20, 31]:

$$E = \omega^{hb} E^{hb} + \omega^{hp} E^{hp} + \omega^{stiff} E^{stiff} + \omega^{elec} E^{elec} =$$
$$= \omega^{hb} \sum_i \sum_j E^{hb}_{i,j} + \omega^{hp} \sum_i \sum_j E^{hp}_{i,j} + \omega^{stiff} \sum_i E^{stiff}_i + \omega^{elec} \sum_i \sum_j E^{elec}_{i,j}$$

where $E^{hb}$ is the energy associated with backbone hydrogen bonds; $E^{elec}$ is the energy associated to electrostatic interactions; and $E^{hp}$ is the energy associated with hydrophobic interactions. A "stiffness term", $E_i^{stiff}$, has also been included. While specific details and tests of validity are extensively discussed in previous work [10, 20, 31], here we report only the main features of the terms above.

Any pair of residues $i$ and $j$ (where $j = i + 2$ or $j \geq i + 4$) may form a hydrogen bond interaction, $E^{hb}$. This interaction is calculated in two steps; first, three geometrical conditions are checked (the length of the possible hydrogen bond between beads, the orientation between bond vectors, and the relative orientation between those vectors and the possible hydrogen bond). Second, a step-wise potential is applied if the values associated to such conditions fall within certain ranges. The acceptance ranges as well as the interaction strength depend on the specific nature of the hydrogen bond (local/helical or non-local/$\beta$-type), see Ref. [31] for details.

The hydrophobic interaction, $E^{hp}$ is modeled by a pair-wise 10-12 Lennard-Jones interaction. Hydrophobic and polar centers of interaction can be distinguished by the particular parameters in the Lennard-Jones potential [10]. Amino acids have been divided into the two categories above following Ref. [32]. A sequence-independent stiffness term, $E^{stiff}$, is included to equally favor helical and extended conformations [10, 32] and has a similar role as the torsional interaction term in atomistic simulations. The stiffness term is calculated for each residue, $i$, and defined by a sinusoidal function that depends on the bond angle between three consecutive beads, $\phi$. Minima of the potential are at $\phi_i = \pm\pi/3$ (i.e. the most probable angle of $\alpha$-helices is then well described) and $\phi_i = \pm\pi$ (i.e. the most probable angle in extended configurations such as $\beta$-sheets).

Up to this point, the effect of the pH is not included, as we have proved that the interactions described above are not significantly affected by pH changes [20]. Electrostatic interactions, however, are sensitive to changes in pH. Using the principle of "consistency across the scales", we have designed an "effective" and physically sound electrostatic term, $E^{elec}$, which is modeled by a Yukawa potential [33]. Appropriate tests on a rather large class of peptides can be found in Refs. [20, 21].

The system pH itself is modeled via discrete protonation states, which are allowed to fluctuate during the simulation according to a Monte Carlo scheme [20]. In each Monte Carlo "pH move", a protonable site is randomly selected and its protonation state is changed.

The new configurations is accepted or rejected according to the detailed balance condition, which depends on the "sensitivity" of the specific amino acid towards protonation (ie. the corresponding equilibrium acidity constant) and on the particular pH value.

The balance among the contributions is defined in terms of the following weighting factors: $\omega^{hb} = 9.5$, $\omega^{hp} = 6.5$, $\omega^{stiff} = 7.0$ and $\omega^{elec} = 12.0$. Energetic units are referred to a certain reference temperature $T_{ref}$ and energy $E_{ref} = k_B T_{ref}$; they can be directly related to states typical of experiments. Our simulations were performed according to these reduced and adimensional units, that is $T^* = T_{real}/T_{ref}$ and $E^* = E_{real}/k_B T_{ref}$.

### B. Simulation details

Equilibrium simulations were carried out using a Replica Exchange Monte Carlo (REMC) in-house software, coded in FORTRAN90 and parallelized with OpenMP for higher performance; each of these simulations presents 32 different temperatures. Simulations start from a completely extended conformation for each chain and consists of $8 \cdot 10^6$ Monte Carlo cycles at every temperature after $5 \cdot 10^6$ equilibration cycles (enough to guarantee convergence). In each cycle, every bead of the system is subjected to a trial Monte Carlo move. In order to sample the conformational space as efficiently as possible, we have implemented individual bead movements as well as whole chain movements. See Ref. [31] for details. For each system, at least three independent runs have been performed.

We used Kinetic Monte Carlo (KMC) to extract qualitative dynamic information [34]. In this technique it is assumed that the main process (aggregation, in this case) is achieved in a large number of much smaller steps (Monte Carlo cycles), proportional to the real timescale of the process. As this is true only if small amplitude movements are allowed in the code, larger amplitude Monte Carlo movements (present in the REMC version) were removed from the in-house FORTRAN90 code. The evolution of the KMC system can thus be compared to the real one. In our implementation, random initial configurations are run at a temperature below the transition one, registering the evolution of the system until it reaches the stopping point. Kinetic aggregation simulations were performed at a temperature $T^* = 0.9 \cdot T_m^*$ and stopped when the system energy reached the average energy at the equilibrium transition temperature; in all cases this is shortly after the aggregation nucleus is formed. In order to achieve proper statistics, 500 independent aggregation events have been performed for each system.

In all cases the simulated systems are composed of twenty-four peptides in a simulation box of 21 nm, using periodic boundary conditions to mimic bulk behavior. pH changes are acknowledged through changes in the effective bead-bead interactions while the solvent has kept its implicit representation, as described above.

Simulation results were analyzed using standard protocols, mostly using in-house software. Heat capacity curves were calculated at each simulated temperature by means of thermal fluctuations of the total energy. The number of intermolecular hydrogen bonds and hydrophobic and electrostatics contacts were computed based on the intermolecular contribution of each type of interaction to the overall energy; two beads were considered in contact if their energy was at least 10% of its maximum strength. Aggregate sizes were determined by counting the number of peptides whose centers of mass were closer than 1 nm (the overall results showed minimal differences within a 30% variation of this interchain distance). VMD [35] was used to produce pictures.

## III. SIMULATION RESULTS

This article studies the effect of pH and sequence on peptide aggregation in a systematic way. For this purpose we used realistic systems sizes, ie. larger than the critical nucleus size. This critical size is a function of the peptide sequence and its length and it is usually set (by both experimental and computational means) between four and ten polypeptide chains [11, 28–30]. We used twenty-four peptides per simulation box in Replica Exchange Monte Carlo and Kinetic Monte Carlo pH-dependent simulations; it is the first time that this large scale approach has been used to take into account the effect of pH in the aggregation process.

We took the *de novo* sequence KTVIIE as starting point, designed by López de la Paz and co-workers as intrinsically amyloidogenic [23, 24]. In the first part of this study we explored the effect of pH in aggregation, as it had been reported that peptides present pH-sensitive aggregation propensities, probably linked to a difference in their net charge [23, 36]. Then, we simulated the KTVIIE peptide with capped ends at three different pH conditions: acidic (pH=2.6), neutral (pH=7.4) and basic (pH=12.5). According to experimental evidence, the three systems aggregate, although these aggregates present slight morphological differences [23, 24].

The second part of this work deals with two related sequences, KTWEFE and PRVIIR. We used them to analyze the role in aggregation of the amyloidogenic core of the original sequence, VII. In the first derived sequence, this triad was mutated to WEF, which is also amyloidogenic but aggregates at a much slower rate [24]. The last sequence, PRVIIR, surrounds the amyloid-forming core by amino acids that prevent aggregation [23, 37]. Peptide termini were set as capped/free to achieve in all cases a peptide net charge of ±1 in an acidic environment (pH 2.6).

### A. Effect of pH in the aggregation of KTVIIE

In this case we evaluated the aggregation characteristics of the KTVIIE peptide at acidic, neutral and basic pH conditions. We analyzed the thermodynamic stability of the KTVIIE peptide system in equilibrium simulations at different temperatures. The results are summarized in the heat capacity curves versus temperature shown in Figure 1A. The curves corresponding to the three tested environments present a peak at temperature $T_m^* = 2.35$, indicative of an energetic transition: the system is forming aggregates at temperatures below $T_m^*$, which dissociate at higher temperatures. This transition is found at a similar temperature at all pH values, which implies that they all aggregate in similar circumstances, in agreement with experimental evidence [23].

However, the height of the heat capacity peak differs, especially at low pH, suggesting microscopic differences among the three systems. We have taken each of them separately and analyzed their intermolecular contacts at a temperature slightly below the transition one ($T^* = 0.9\ T_m^*$). In Figures 1B, 1C and 1D we show the distribution of intermolecular electrostatic interactions, hydrophobic interactions and hydrogen bonds, respectively. Hydrophobic contacts and hydrogen bonds (Figures 1C and 1D) exhibit a comparable distribution at the three pH values, which reflect the similarity of the overall aggregated structure in all cases. These aggregates are formed by hydrogen bonded $\beta$-type structures, which are associated by hydrophobic interactions (see structural discussion referring to Figure 2). Electrostatic interactions, shown in Figure 1B, present a higher variability upon pH, as they are directly affected by the media acidity. In neutral conditions positive and negative charges coexist (according to our simplified model description), which results in a slightly higher number of electrostatic contacts. Basic and especially acidic pH conditions shift the distribution maxima towards lower values. This shift might be due to the fact that in these conditions peptides present an effective net charge, either positive (at pH 2.6) or negative (at pH 12.5), and there are just a few interaction centers of opposite charge to interact with. These residual electrostatic interactions, however, remain present even above the transition temperature ($T^* = 2.50$, data not shown), which suggests that some degree of electrostatic attraction persist at higher temperatures.

Our thermodynamics results indicate that KTVIIE is indeed prone to form aggregates in all sorts of conditions, as experimentally reported [23, 24]. Therefore, a very simple model like ours is sensitive to sequence and it can qualitatively grasp intrinsic pH-dependent effects. Therefore, this "toy" pH-dependent potential can be used in realistic applications while keeping its simplicity and inexpensive computational cost.

Regarding the aggregation dynamics of this peptide, we carried out KMC calculations for the three different pH conditions used in the equilibrium simulations. This approach, quite low in resolution but very suited for the kind of coarse-grained potential used in this work, allows to use large size systems at a limited computational cost. Therefore, it was possible to register hundreds of independent aggregation events for each particular condition, permitting a proper statistical analysis of KTVIIE aggregation kinetics.

The structural description of the aggregation dynamics is quite similar in all our simulation runs, as illustrated in Figure 2. The Figure shows the most relevant stages in a typical aggregation event and shows the key aspects regarding how aggregation takes place. This particular example was taken from one single run of KTVIIE at pH 2.6, although just minor differences were found in neutral and basic environmental conditions. Snapshots A to D show the evolution of the whole system at different times $t$, relative to the total aggregation time (denoted by $\tau$). Simulations are started from random configurations (see Figure 2A), which are allowed to diffuse freely according to our KMC scheme. The magnified image in the Figure reflects the fact that peptides adopt random configurations with residual intramolecular hydrogen bonds in some peptide chains.

As time evolves, sporadic hydrophobic interactions are formed among peptides; they become more frequent and robust around $t = 0.5\,\tau$ (see Figure 2B). The enlarged image shows three independent peptides (represented in different colors for the sake of clarity) agglutinated into an unspecific cluster. It is mostly stabilized by hydrophobic interactions and electrostatics, with null or little presence of hydrogen bonds in the overall structure. These clusters are transient, ie. they rapidly form and dissociate. However, the large size of the simulated system allows us to identify the formation of many similar clusters (usually composed by 3-5 peptide chains) at this stage of the simulation (as can be observed in a close inspection of the global image in Figure 2B). The formation of this *hydrophobic core* is, then, identified as a key step in the aggregation process, in agreement with literature [38].

The hydrophobic clusters rearrange at $t = 0.75\ \tau$, as observed in Figure 2C. At this point there is a shift in the stabilizing interactions, from mostly hydrophobic contacts to hydrogen bonds. Peptide chains acquire a more elongated form, which maximizes the number of intermolecular hydrogen bonds and form a $\beta$-sheet type structure; its hallmark parallel distribution can be easily recognized in the magnified cartoon in Figure 2C. The *hydrogen-bonded core* structure acts like a "seed" that promotes the growth of the aggregate (fibril), as shown in Figure 2D for time $t = \tau$. This final structure is similar to the ones observed in REMC equilibrium simulations, proving that convergence was reached. Note also that a purely hydrogen bonded configuration would result in a highly exposed and unstable hydrophobic core; this is avoided by adding more peptides to the structure, usually through hydrophobic interactions (see the peptide colored in magenta in Figure 2D).

Thanks to the KMC technique, a large number of aggregation events were registered in each particular case (see Section II for details), resulting in meaningful statistics about aggregation. We calculated the relative number aggregation events ($N/N_0$) as a function of time, plotted in Figure 3. This representation provides information about the kinetics law that drive aggregation according to our model; in addition, it illustrates the differences between the tested conditions. In Figure 3, the three simulated systems (acidic, neutral and basic) show a linear trend, consistent with a first order kinetics. This finding is in agreement with data from other aggregation studies [39] and suggests that it is a common feature in peptide aggregation.

The fitting details from our simulations are shown in Table I. Results in this Table indicate that the aggregation rate is a pH-dependent property for the KTVIIE peptide according to our model. The rate constant is larger at neutral pH; this suggests that neutral pH conditions not only averages out the overall peptide net charge but also promotes intermolecular interactions, as more beads with opposite charges are present in the system. This also suggests that, at least according to our simplified interaction scheme, systems with non-zero net charge do not necessarily aggregate faster.

So far, we have performed a rigorous study of the aggregation conditions for the KTVIIE peptide. We have shown that a simple but carefully designed one-bead coarse-grained model such as ours is able to grasp the essence of peptide aggregation from both a kinetic and a thermodynamic point of view; this kind of approach, then, can provide valuable insight to interpret and guide experimental studies. Besides, the effect of pH is correctly tackled by the coupling of our coarse-grained model with a pH-dependent algorithm, highlighting the impact of the re-distribution of the system electrostatics upon acidity changes.

### B. Sequence-dependent effects on aggregation at low pH

The KTVIIE sequence is unique in its kind, in the sense that its aggregation propensity is so high that some degree of aggregation is expected in nearly any condition. However, modifications in its sequence can lead to noticeable differences in the aggregation response [24]. We investigated this aspect by simulating two additional sequences that focus on the role of the amyloidogenic core VII. The first one, KTWEFE, presents the WEF motif instead of VII; WEF is thought to slow down aggregation. Our second variation is the sequence PRVIIR, which presents one proline and two positively charged amino acids surrounding the aggregation motif VII. The former has been identified as a $\beta$-sheet breaker amino acid [37]; the latter are thought to destabilize aggregates due to an increase in the local peptide charge [23].

Regarding the first alternative sequence, experimental evidence at low pH conditions states that the KTWEFE peptide has a much slower kinetic response [24], although it still aggregates at long timescales (ie. in the thermodynamic regime). Using a model as simple as ours, all the chemical subtleties regarding these mutations cannot be grasped. However, our model is able to detect that the new sequence is more polar and has more charged interaction centers; these features should be clearly reflected in our results. This hypothesis was tested by simulating the KTWEFE peptide at pH 2.6, using REMC and KMC to extract thermodynamic and dynamic information about the effect that the new sequence has in aggregation.

Our equilibrium simulations showed that KTWEFE is able to form aggregates at low pH, presenting a peak in the heat capacity curve at $T_m^* = 2.50$, similarly to the KTVIIE curves in Figure 1 (data not shown). The slight increase in the transition temperature reflects the fact that KTWEFE may form more stable aggregates than the original sequence. We compared the aggregate sizes at $T^* = 0.9\, T_m^*$ for the two sequences, shown in blue and red in Figure 4. KTWEFE aggregates are slightly larger in size. Regarding their structure, they are similar to KTVIIE aggregates, with a mild increase in the average number of hydrogen bonds due to the incorporation of more peptide chains into the hydrogen bonded aggregated cluster (similar to Figure 2D). These evidences support the fact that KTWEFE aggregates are thermodynamically stable, like those in their parent sequence KTVIIE.

In the case of the KMC dynamic information, first order kinetics are again fulfilled, as shown in Table I by a fitting correlation coefficient very close to one ($R^2 = 0.99$). The main difference compared to KTVIIE was found in the rate constant, which is in this case an order of magnitude smaller than in KTVIIE. A smaller rate constant matches the experimental evidence of slower aggregation kinetics for this sequence compared to the original one [24]. It also highlights the fact that a thermodynamically possible event may or may not occur in dynamically driven conditions such as a time-dependent experiment, as the rate constant may be largely affected. It is worth mentioning that this difference has been spotted with a one bead coarse-grained model. This type of models, by definition, lack a side chain explicit description: only a careful design of the bead to bead interactions (such as the present model) can make up for sequence-dependent details, which could be easily lost otherwise.

The original sequence KTVIIE has also been modified in the literature to introduce amino acids that are thought to prevent aggregation [23, 37]. One of these alternative sequences is PRVIIR, which surrounds the aggregation key motif VII by so-called $\beta$-sheet breaker amino acids [24, 37]. We simulated the new sequence in equilibrium conditions using a similar temperature range as in all the other cases. To our surprise, we did not observe a shift of the heat capacity peak towards lower temperatures, but its complete disappearance, which means that PRVIIR does not aggregate at any of the simulated temperatures. A numerical analysis of the aggregate sizes at low temperatures shows, in fact, no presence of intermolecular structures, as can be observed in the yellow histogram of Figure 4. For this reason, no kinetic study was performed in this case. PRVIIR results illustrate the importance of mutations in aggregation, as they can completely prevent the formation of intermolecular structures.

## IV. CONCLUSIONS

Peptides have been extensively used as test systems to study fundamental aspects of protein aggregation, such as the role of each amino acid in promoting/disrupting polymerization, the effect of net charge in the overall process or the impact that environment acidity may have in the formation of aggregates. One of the most common peptides for this kind of studies is the *de novo* sequence KTVIIE, broadly investigated by experiments and simulations [23–27].

Computational tools play a preponderant role in aggregation studies, as they can provide microscopic detail as well as fundamental information. However, simulation time is still a hard limit for many applications due to the size of the relevant systems. This is especially true in pH-dependent studies, as current atomistic constant-pH methodologies are about an order of magnitude slower than their traditional counterparts. In this work we have applied a recently developed coarse-grained pH-dependent strategy to circumvent this issue. Thanks to its simplicity and low computational cost, we have been able to simulate three different multichain systems at several pH conditions. This study has been carried out using REMC and KMC to extract both thermodynamic and kinetic information.

We have found, in accordance to the literature, that KTVIIE is intrinsically amyloidogenic in all pH conditions, presenting slight but noticeable differences depending on the particular media acidity. The sequence variation KTWEFE, also amyloidogenic, aggregates at much slower rates. The sequence PRVIIR, which presents the amyloidogenic motif VII but is surrounded by aggregation defender amino acids, does not aggregate in our simulations. These findings highlight the subtleties that drive protein aggregation, so further fundamental studies are needed. Furthermore, our coarse-grained pH-dependent methodology has proved to be a valuable tool to study pH-dependent processes in a fast, reliable and computationally affordable way, which makes it suitable to interpret and guide experiments in the future.

**Acknowledgment**

This research has been funded by Deutsche Forschungsgemeinschaft (DFG) through grant CRC 1114. ME acknowledges funding from the Australian National Health and Medical Research Council and the Australian Research Council. The authors would also like to thank VPAC and the Victorian Life Science Initiative for the use of their computational resources.[1] Hanns-Christian Mahler, Wolfgang Friess, Ulla Grauschopf, and Sylvia Kiese. Protein aggregation: Pathways, induction factors and analysis. *J. Pharm. Sci.*, 98(9):2909–2934, 2009.

[2] Adriano Aguzzi and Tracy O'Connor. Protein aggregation diseases: pathogenicity and therapeutic perspectives. *Nat. Rev. Drug Discovery*, 9(3):237–248, 2010.

[3] Sven Frokjaer and Daniel E Otzen. Protein drug stability: a formulation challenge. *Nat. Rev. Drug Discovery*, 4(4):298–306, 2005.

[4] Jeppe T. Pedersen and Niels H. H. Heegaard. Analysis of protein aggregation in neurodegenerative disease. *Anal. Chem.*, 85(9):4215–4227, 2013.

[5] Alexandra K Marr, William J Gooderham, and Robert EW Hancock. Antibacterial peptides for therapeutic use: obstacles and realistic outlook. *Curr. Opin. Pharma.*, 6(5):468–472, 2006.

[6] Justin A. Lemkul and David R. Bevan. The role of molecular simulations in the development of inhibitors of amyloid $\beta$-peptide aggregation for the treatment of alzheimers disease. *ACS Chem. Neurosci.*, 3(11):845–856, 2012.

[7] Chun Wu, Justin Scott, and Joan-Emma Shea. Binding of congo red to amyloid protofibrils of the alzheimer a$\beta$ 9–40 peptide probed by molecular dynamics simulations. *Biophys. J.*, 103(3):550–557, 2012.

[8] Luca Larini and Joan-Emma Shea. Role of $\beta$-hairpin formation in aggregation: The self-assembly of the amyloid-$\beta$ (25–35) peptide. *Biophys. J.*, 103(3):576–586, 2012.

[9] Martín Carballo-Pacheco, Ahmed E. Ismail, and Birgit Strodel. Oligomer formation of toxic and functional amyloid peptides studied with atomistic simulations. *J. Phys. Chem. B*, 119(30):9696–9705, 2015.

[10] Marta Enciso and Antonio Rey. Simple model for the simulation of peptide folding and aggregation with different sequences. *J. Chem. Phys.*, 136(21):215103, 2012.

[11] Nicolas Lux Fawzi, Yuka Okabe, Eng-Hui Yap, and Teresa Head-Gordon. Determining the critical nucleus and mechanism of fibril elongation of the alzheimers a$\beta$ 1–40 peptide. *J. Mol. Biol.*, 365(2):535–550, 2007.

[12] J Iñaki Guijarro, Craig J Morton, Kevin W Plaxco, Iain D Campbell, and Christopher M
14

Dobson. Folding kinetics of the sh3 domain of pi3 kinase by real-time nmr combined with optical spectroscopy. *J. Mol. Biol.*, 276(3):657–667, 1998.

[13] Fernanda G. De Felice, Marcelo N. N. Vieira, M. Nazareth L. Meirelles, Ludmilla A. Morozova-Roche, Christopher M. Dobson, and Srgio T. Ferreira. Formation of amyloid aggregates from human lysozyme and its disease-associated variants using hydrostatic pressure. *FASEB J.*, 2004.

[14] Anthony W. Fitzpatrick, Tuomas P. J. Knowles, Christopher A. Waudby, Michele Vendruscolo, and Christopher M. Dobson. Inversion of the balance between hydrophobic and hydrogen bonding interactions in protein folding and aggregation. *PLoS Comput Biol*, 7(10):e1002169, 10 2011.

[15] Michael H Hecht. De novo design of beta-sheet proteins. *Proc. Natl. Acad. Sci.*, 91(19):8729–8730, 1994.

[16] Mapping of hippocampal pH and neurochemicals from in vivo multi-voxel 31P study in healthy normal young male/female, mild cognitive impairment, and Alzheimer's disease. *J Alzheimers Dis.*, 31:S75, 2012.

[17] Wei Wang. Protein aggregation and its inhibition in biopharmaceutics. *Intl. J. Pharm.*, 289(1):1–30, 2005.

[18] John Mongan and David A. Case. Biomolecular simulations at constant 157-163pH. *Curr. Opin. Struct. Biol.*, 15(2):157–163, 2005.

[19] Miguel Machuqueiro and Antnio M. Baptista. Acidic range titration of hewl using a constant-ph molecular dynamics method. *Proteins*, 72(1):289–298, 2008.

[20] Marta Enciso, Christof Schütte, and Luigi Delle Site. A pH-dependent coarse-grained model for peptides. *Soft Matter*, 9(26):6118–6127, 2013.

[21] Marta Enciso, Christof Schütte, and Luigi Delle Site. pH-dependent response of coiled coils: a coarse-grained molecular simulation study. *Mol. Phys.*, 111(22-23):3363–3371, 2013.

[22] Matej Praprotnik, Luigi DelleSite, and Kremer. Multiscale simulation of soft matter: From scale bridging to adaptive resolution. *Annu. Rev. Phys. Chem.*, 59:545–571, 2008.

[23] Manuela López de la Paz, Kenneth Goldie, Jesús Zurdo, Emmanuel Lacroix, Christopher M Dobson, Andreas Hoenger, and Luis Serrano. De novo designed peptide-based amyloid fibrils. *Proc. Natl. Acad. Sci.*, 99(25):16052–16057, 2002.

[24] Manuela López de la Paz and Luis Serrano. Sequence determinants of amyloid fibril formation.

*Proc. Natl. Acad. Sci.*, 101(1):87–92, 2004.

[25] Joohyun Jeon and M.Scott Shell. Charge effects on the fibril-forming peptide ktviie: A two-dimensional replica exchange simulation study. *Biophys. J.*, 102(8):1952 – 1960, 2012.

[26] Wonmuk Hwang, Shuguang Zhang, Roger D. Kamm, and Martin Karplus. Kinetic control of dimer structure formation in amyloid fibrillogenesis. *Proc. Natl. Acad. Sci.*, 101(35):12916–12921, 2004.

[27] Juyong Lee, Benjamin T. Miller, Ana Damjanovi, and Bernard R. Brooks. Enhancing constant-ph simulation in explicit solvent with a two-dimensional replica exchange method. *J. Chem. Theor. Comput.*, 11(6):2560–2574, 2015.

[28] Ute F Röhrig, Alessandro Laio, Nazario Tantalo, Michele Parrinello, and Roberto Petronzio. Stability and structure of oligomers of the alzheimer peptide a$\beta$ 16–22: from the dimer to the 32-mer. *Biophys J.*, 91(9):3217–3229, 2006.

[29] Yan Lu, Philippe Derreumaux, Zhi Guo, Normand Mousseau, and Guanghong Wei. Thermodynamics and dynamics of amyloid peptide oligomerization are sequence dependent. *Proteins*, 75(4):954–963, 2009.

[30] Karunakar Kar, Murali Jayaraman, Bankanidhi Sahoo, Ravindra Kodali, and Ronald Wetzel. Critical nucleus size for disease-related polyglutamine aggregation is repeat-length dependent. *Nature Struct. Mol. Biol.*, 18(3):328–336, 2011.

[31] Marta Enciso and Antonio Rey. A refined hydrogen bond potential for flexible protein models. *J. Chem. Phys.*, 132:235102 1–10, 2010.

[32] Scott Brown, Nicolas J Fawzi, and Teresa Head-Gordon. Coarse-grained sequences for protein folding and design. *Proc. Natl. Acad. Sci.*, 100(19):10712–10717, 2003.

[33] Sergei Izvekov, Jessica M. J. Swanson, and Gregory A. Voth. Coarse-graining in interaction space: A systematic approach for replacing long-range electrostatics with short-range potentials. *J. Phys. Chem. B*, 112(15):4711–4724, 2008.

[34] Rui DM Travasso, Patrícia FN Faísca, and Antonio Rey. The protein folding transition state: Insights from kinetics and thermodynamics. *J. Chem. Phys.*, 133(12):125102, 2010.

[35] William Humphrey, Andrew Dalke, and Klaus Schulten. Vmd: visual molecular dynamics. *Journal of molecular graphics*, 14(1):33–38, 1996.

[36] Fabrizio Chiti, Martino Calamai, Niccol Taddei, Massimo Stefani, Giampietro Ramponi, and Christopher M. Dobson. Studies of the aggregation of mutant proteins in vitro provide insights

into the genetics of amyloid diseases. *Proc. Natl. Acad. Sci.*, 99(suppl 4):16419–16426, 2002.

[37] Céline Adessi and Claudio Soto. Beta-sheet breaker strategy for the treatment of alzheimer's disease. *Drug Dev. Res.*, 56(2):184–193, 2002.

[38] Nicolae-Viorel Buchete and Gerhard Hummer. Structure and dynamics of parallel $\beta$-sheets, hydrophobic core, and loops in alzheimers a$\beta$ fibrils. *Biophys. J.*, 92(9):3032–3039, 2007.

[39] William P Esler, Evelyn R Stimson, Joseph R Ghilardi, Harry V Vinters, Jonathan P Lee, Patrick W Mantyh, and John E Maggio. In vitro growth of alzheimer's disease $\beta$-amyloid plaques displays first-order kinetics. *Biochemistry*, 35(3):749–757, 1996.

## Tables and Figures

| System | $k$ ($10^{-6}/MCcycles$) | $R^2$ |
|---|---|---|
| KTVIIE, pH 2.6 | 8.95 | 0.98 |
| KTVIIE, pH 7.4 | 22.5 | 0.97 |
| KTVIIE, pH 12.5 | 12.6 | 0.94 |
| KTWEFE, pH 2.6 | 0.47 | 0.99 |

TABLE I: Kinetic information regarding the peptide KTVIIE at three different pH conditions and KTWEFE at low pH. $k$ is the first order rate constant, calculated from the fitting in Figure 3 with correlation coefficient $R^2$.

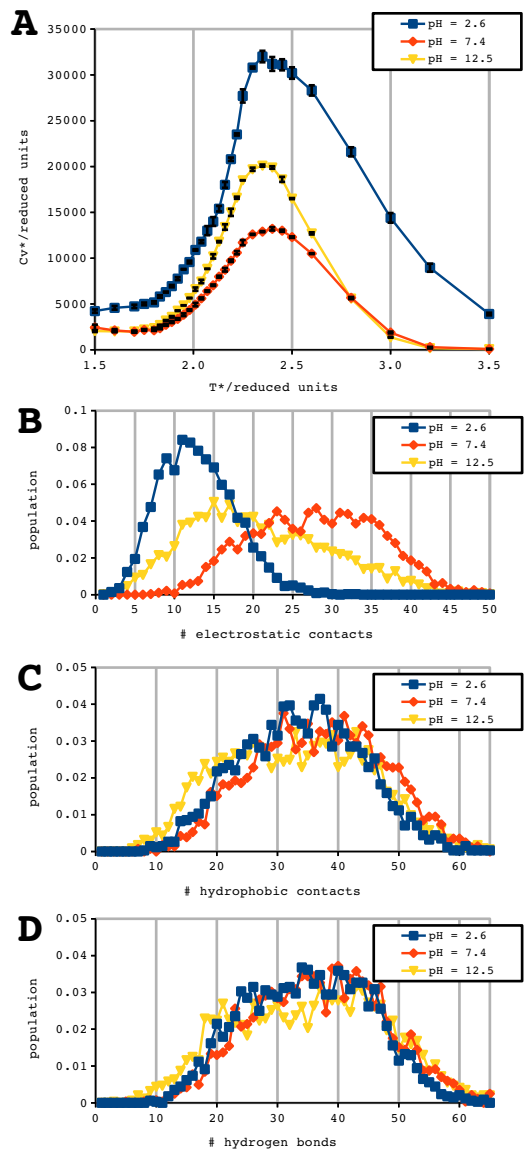

FIG. 1: Thermodynamic and structural information for KTVIIE at pH 2.6 (blue), 7.4 (red) and 12.5 (dark yellow). A) Heat capacity curve versus temperature, in reduced units. Contact histogram for B) electrostatic interactions, C) hydrophobic interactions and D) hydrogen bonds.

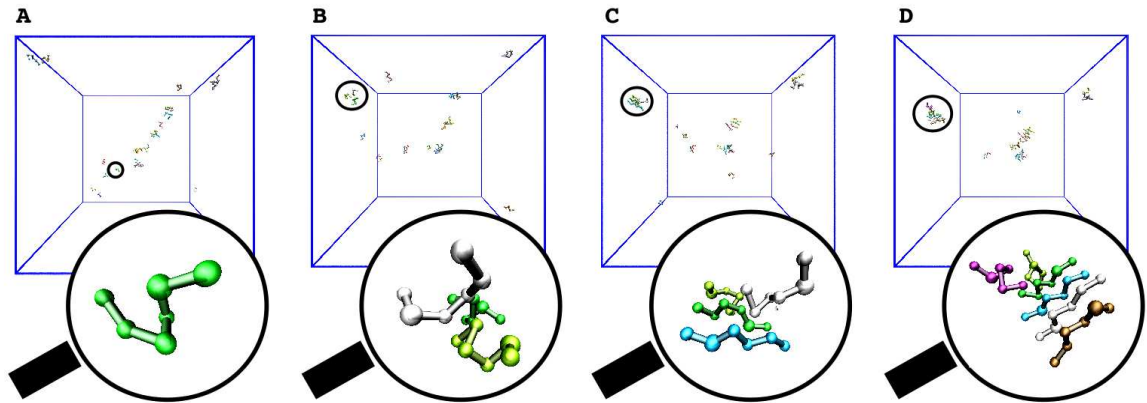

FIG. 2: Stages of the aggregation process for KTVIIE at pH 2.6. A) At $t = 0$ peptides present random configurations. B) At $t = 0.5\ \tau$, peptides interact, forming a hydrophobic collapsed cluster. C) At $t = 0.75\ \tau$, collapsed states rearrange and form hydrogen bonds. D) At $t = \tau$ the hydrogen bonded aggregates can freely grow.

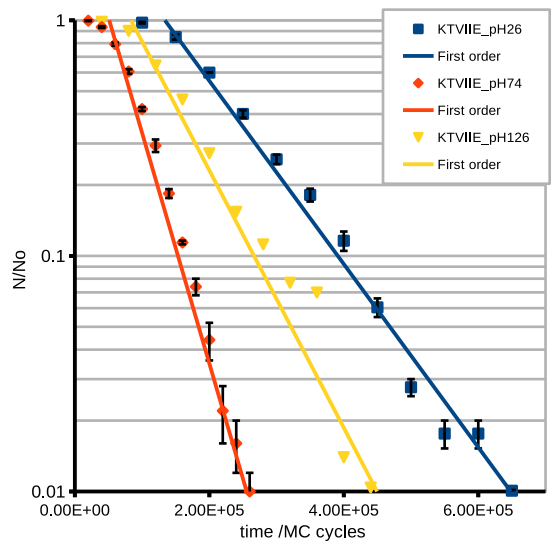

FIG. 3: Relative population of isolated peptides (ie. not forming aggregates) vs time (in number of Monte Carlo cycles) for the sequence KTVIIE at $T_m^* = 2.10$ and two different pH conditions

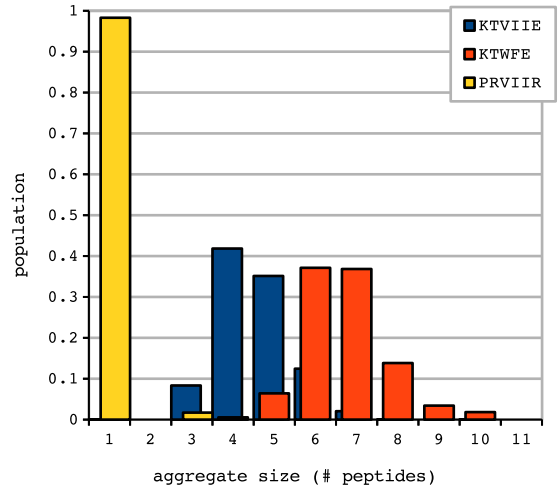

FIG. 4: Aggregate size for the studied sequences a pH=2.6 at $T^* = 0.9\ T_m^*$. Blue: KTVIIE at $T^* = 2.10$; red: KTWEFE at $T^* = 2.30$; yellow: PRVIIR at $T^* = 1.50$ (lowest simulated temperature).